\documentclass{aa}

\usepackage{graphicx}
\usepackage{subfig}
\usepackage[varg]{txfonts}
\usepackage{ngerman}

\begin{document}

   \title{The Discovery of a Low-Luminosity SPIRAL DRAGN.}

   \author{D.~D.~Mulcahy\inst{1}
          \and M.~Y.~Mao\inst{1}
          \and I.~Mitsuishi\inst{2}
          \and A.~M.~M.~ Scaife\inst{1}
          \and A.~O.~Clarke\inst{1}
          \and Y.~Babazaki\inst{2}
          \and H.~Kobayashi\inst{2}
          \and R.~Suganuma\inst{2}
          \and H.~Matsumoto\inst{3}
          \and Y.~Tawara\inst{2}
          }

   \institute{
             Jodrell Bank Centre for Astrophysics, Alan Turing Building, School of Physics and Astronomy,\\ The University of Manchester, Oxford Road, Manchester, M13 9PL, U.K.\\
             \email{david.mulcahy@manchester.ac.uk}
             \and
             Division of Particle and Astrophysical Science, Nagoya University, Furo-cho, Chikusa-ku, Nagoya, Aichi, 464-8602 JAPAN 
             \and 
             Kobayashi-Maskawa Institute for the Origin of Particles and the Universe, Nagoya University, Furo-cho, Chikusa-ku, Nagoya, Aichi, 464-8602 JAPAN
             }

   \date{Received 16 August 2016; accepted 13 September 2016}

  \abstract{Standard galaxy formation models predict that large-scale double-lobed radio sources, known as DRAGNs, will always be hosted by elliptical galaxies. In spite of this, in recent years a small number of spiral galaxies have also been found to host such sources. These so-called \emph{spiral DRAGNs} are still extremely rare, with only $\sim 5$ cases being widely accepted. Here we report on the serendipitous discovery of a new spiral DRAGN in data from  the Giant Metrewave Radio Telescope (GMRT) at 322\,MHz. The host galaxy, MCG+07-47-10, is a face-on late-type Sbc galaxy with distinctive spiral arms and prominent bulge suggesting a high black hole mass. Using WISE infra-red and GALEX UV data we show that this galaxy has a star formation rate of 0.16-0.75 \,M$_{\odot}$\,yr$^{-1}$, and that the radio luminosity is dominated by star-formation. We demonstrate that this spiral DRAGN has similar environmental properties to others of this class, but has a comparatively low radio luminosity of $L_{\rm 1.4\,GHz}$ = 1.12$\times$10$^{22}$ W Hz$^{-1}$, two orders of magnitude smaller than other known spiral DRAGNs. We suggest that this may indicate the existence of a previously unknown low-luminosity population of spiral DRAGNS.}


   \maketitle

\section{Introduction}

Spiral DRAGNs \citep[Double-lobed Radio sources Associated with Galactic Nuclei, ][]{Leahy1993} are spiral galaxies that host large-scale double-lobed radio sources. The existence of such sources is in contradiction to our existing models of galaxy formation \citep[e.g.][]{Hopkins2008}, which predict that DRAGNs should be hosted exclusively by elliptical galaxies. Indeed, until recently,  observations of DRAGNs in the local Universe confirmed this expectation \citep[e.g.][]{Matthews1964, Urry1995,Best2005}.

Elliptical galaxies are formed as a result of mergers, the phenomenology of which also triggers the formation of DRAGNs \citep{Chiaberge2011,Chiaberge2015}. However, a spiral galaxy's structure cannot withstand a major merger. Moreover, morphological transition from spiral to elliptical is thought to be a one-way process, at least in the local Universe. Consequently, the standard galaxy formation model does not predict the existence of spiral DRAGNs.

Nonetheless, a number of spiral DRAGNs have been discovered in recent years \citep[e.g.][]{Ledlow2001,Hota2011,Bagchi2014,Mao2015,Singh2015}. While the first three spiral DRAGN discoveries were serendipitous, \citet{Mao2015} performed the first systematic search for these sources. Using Galaxy Zoo \citep{Lintott2008} morphological classifications were cross-matched with the Faint Images of the Radio Sky at Twenty-Centimeters \citep[FIRST, ][]{Becker1995} and the NRAO VLA Sky-Survey \citep[NVSS, ][]{Condon1998}. In this study, only one spiral DRAGN was found above L$_{\rm 1.4\,GHz}$ = $10^{23}$\,W\,Hz$^{-1}$. 
\citet{Singh2015} performed a similar analysis using the spiral galaxy catalogue of \citet{Meert2015} and reported the identification of four spiral DRAGNs in those data, including one that was previously known and three that were unknown. However, the precise identification of spiral DRAGN hosts remains contentious in the literature and to date the existence of only 5 spiral DRAGNs are widely accepted. 

DRAGNs with spiral hosts may represent a rare phenomenon of elliptical galaxies transitioning back into spirals through accretion of gas and stars, perhaps from a companion. A key question is whether spiral DRAGNs are a result of non-standard physical properties, a result of their environment, or perhaps a combination of their nature and nurture. Studying spiral DRAGNs, as well as establishing their numbers more exactly, is vital in order to reconcile their role in standard galaxy formation theories.

In this Letter, we present the discovery of a new spiral DRAGN at 325\,MHz with the Giant Meterwave Radio Telescope \citep[GMRT; ][]{Swarup1990}. In Sect.~2 we outline the data processing and imaging steps. In Sect.~3, we present the discovery of this new spiral DRAGN with a description of both its radio morphology and that of the host galaxy, as understood from available multi-wavelength data. In Sect.~4, using Infra-red and UV data we demonstrate that the host galaxy is star-forming and we compare its star formation rate to other spiral DRAGNs.
Finally, in Sect.~5 we discuss the nature of this object and state our conclusions. In this work we assume a $\Lambda$CDM cosmology with $H_0 = 69.6\, \mathrm{km s^{-1} Mpc^{-1}}$, ${\Omega_\mathrm{m}} = 0.286$ and ${\Omega_\Lambda} = 0.714$ \citep{Bennett2014}, which we use for the calculation of distance, luminosity and star formation rate. 
At a redshift of z = 0.017, these values result in a conversion of  0.348\,kpc/$\arcsec$. All uncertainties are quoted at 1\,$\sigma$.

\section{Observations and Data Reduction}

Observations of the galaxy groups NGC7618 \& UGC\,12491 were performed in full synthesis mode with the Giant Meterwave Radio Telescope (GMRT) at 325\,MHz. 
The GMRT is a full aperture synthesis telescope located near Pune, India \citep{Swarup1990}. 
It consists of 30 steerable dishes that are 45\,m in diameter, with a longest interferometric baseline of 25.5\,km, and 14 additional antennas located in a central 1\,sq.\,km, to provide dense \emph{uv} coverage at short spacings. The telescope operates at six frequencies: 150, 230, 325, 610, and 1420\,MHz. In this work the 325\,MHz receiver was utilised. The Full Width Half Maximum (FWHM) of the GMRT primary beam at 325\,MHz is approximately 81$\arcmin\pm4\arcmin$. 
  
The aforementioned observation took place on 16th August 2014 under project code 25.059 (PI Mitsuishi). The observation was a single pointing with a phase center of (J2000) \,23$^\mathrm{h}$ 19$^\mathrm{m}$ 19$^\mathrm{s}$ +42$\degr$ 54$\arcmin$ 50$\arcsec$. The bright radio source 3C48 was used as a flux calibrator, tied to the \cite{Scaife2012} flux density scale. 

The dataset was reduced using the Source Peeling and Atmospheric Modelling (SPAM) software \citep{Intema2009}. The SPAM software performed an initial phase calibration and astrometry correction using a sky model derived from the NVSS \citep[]{Condon1998}, followed by three rounds of direction-independent phase-only self-calibration.
Following this, SPAM performed facet-based direction-dependent calibration on strong sources within the primary beam FWHM, and utilised the solutions from this calibration to fit a global ionospheric model.
During this calibration process, various automated flagging routines were used between cycles of imaging and self-calibration in order to reduce residual RFI and clip statistical outliers.

The final primary beam corrected image towards MCG+07-47-10 is shown in Fig.~\ref{fig:radioimage} and has a resolution of 9.4 $\times$ 8.3$\arcsec$. Images were made using an AIPS ROBUST value of -1.0. This parameter value was selected to achieve an optimal compromise between sensitivity and resolution. More details of these observations and data reduction can be found in Mitsuishi et al., in prep. 

In order to isolate the integrated flux density of the separate source components (radio lobes and core) the source finding software, the Python Blob Detection and Source Measurement software, (PYBDSM\footnote{https://dl.dropboxusercontent.com/u/1948170/html/index.html}) was first used to identify confusing compact sources.
To do this, a local rms noise value ($\sigma$) was calculated in order to identify all significant peaks above a threshold of 7$\sigma$. Following standard practice, PYBDSM then forms islands of contiguous emission down to a threshold of 5$\sigma$ and an elliptical Gaussian is fitted to each island, of which 625 were found in the entire field. Each Gaussian was then subtracted from the image and the residual data used to determine the integrated flux of the radio lobes. The measured integrated fluxes are presented in Table.~\ref{Tab:locationsandflux}. The figures shown in this Letter are not the source subtracted images.

\section{Discovery of spiral DRAGN -- MCG+07-47-10}

\begin{figure}[h!]
\centering
\includegraphics[width=0.35\textwidth,angle=-90]{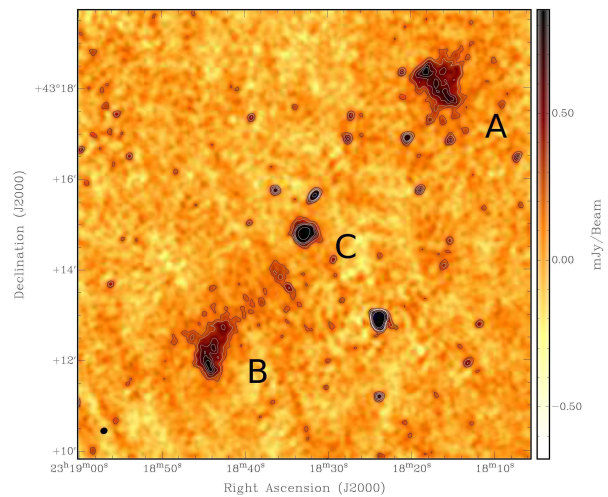}
\caption{GMRT data at 322\,MHz. The local rms noise around the source is approximately $\sigma_{\rm rms} = 68\,\mu$Jy/beam. Contours are shown at 3, 5, 8, 10, 12, 24 \& 48 $\sigma_{\rm rms}$. The GMRT synthesized beam has dimensions of 9.4$\times$8.3$\arcsec$ and is shown in the bottom left corner as a filled-in ellipse.}
\label{fig:radioimage}
\end{figure}

\begin{figure*}
    \centering
\subfloat{{\includegraphics[width=0.3\textwidth]{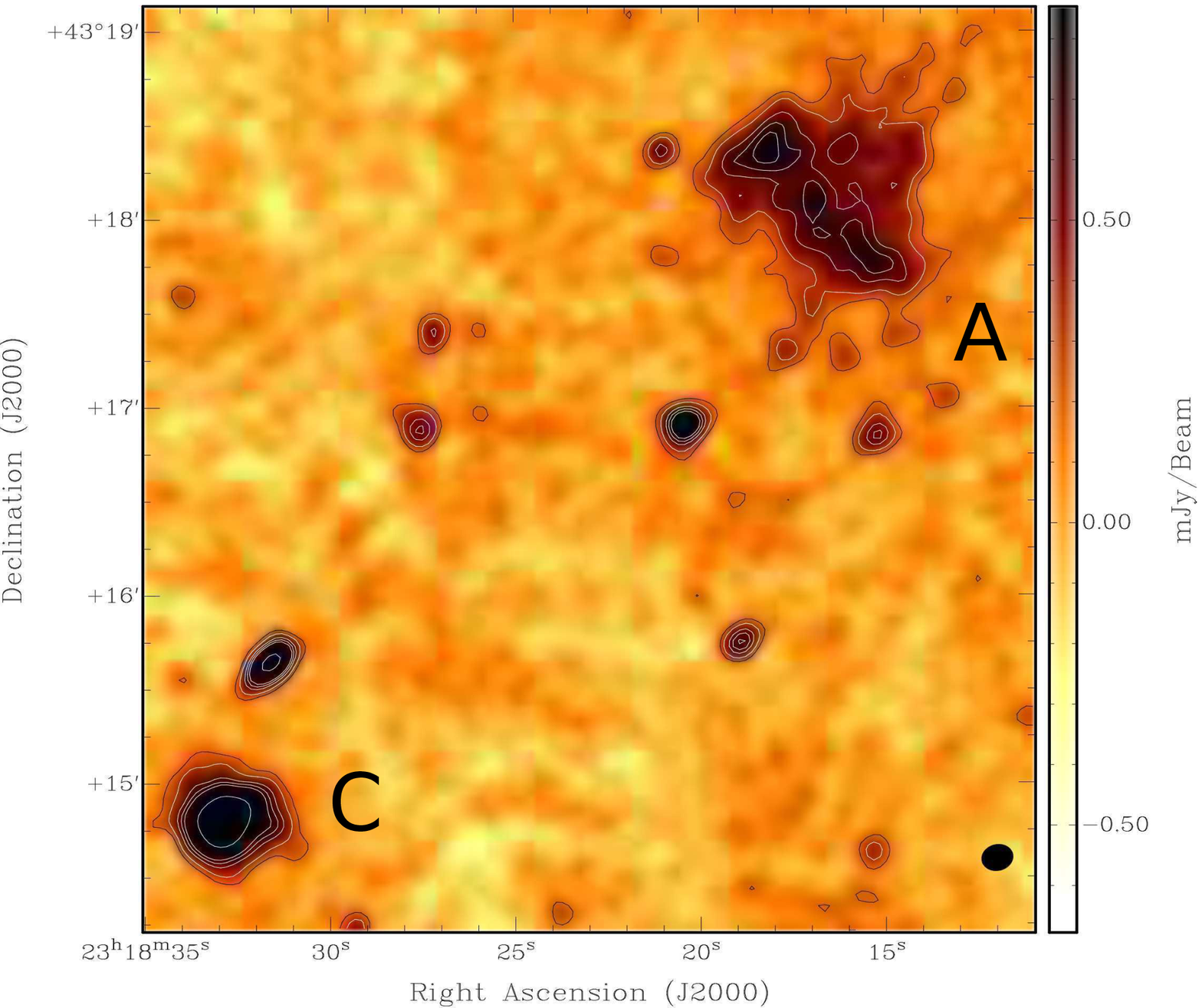} }}%
\subfloat{{\includegraphics[width=0.3\textwidth]{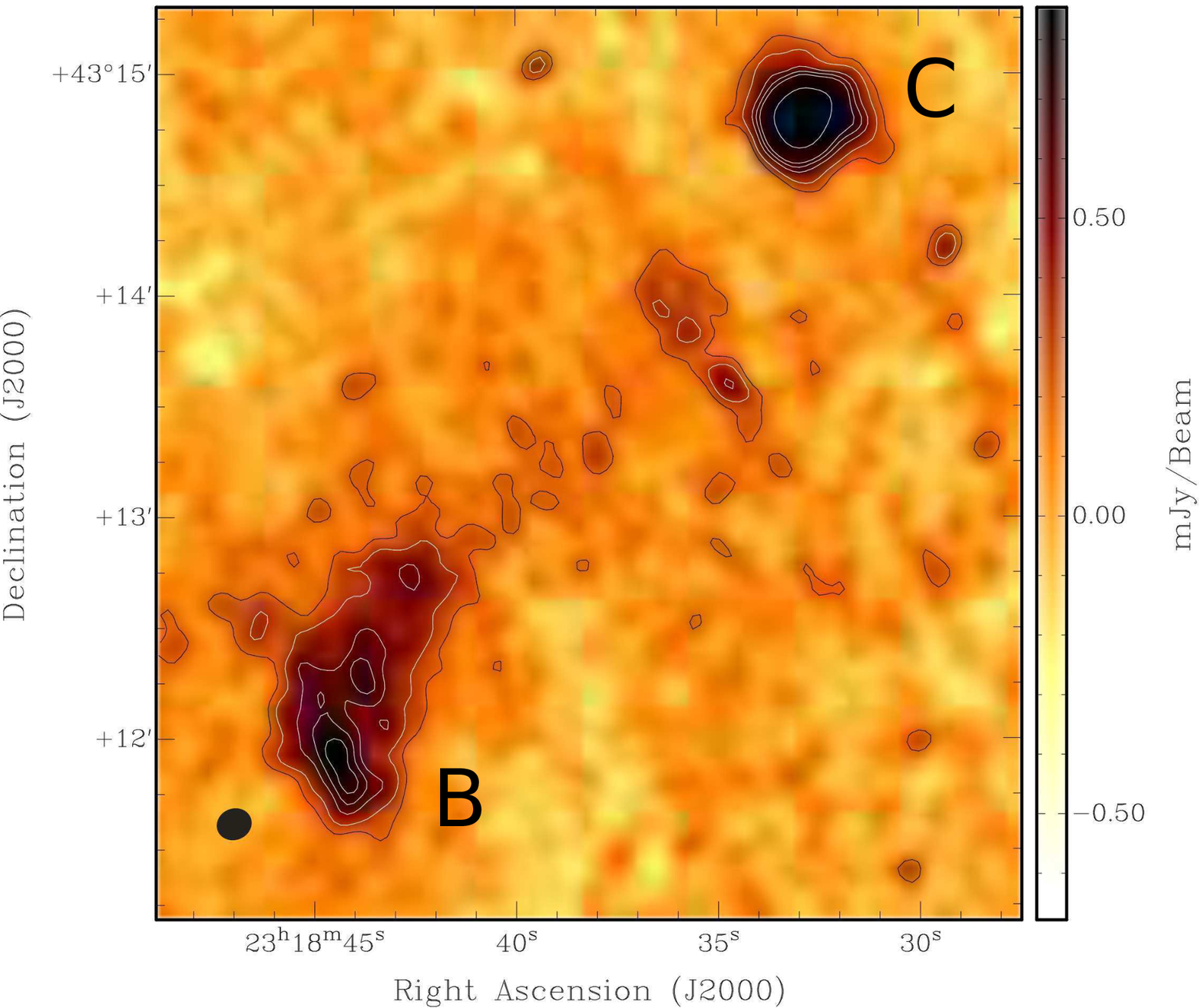} }}%
    \caption{Left: Zoomed in image of the northern lobe and host galaxy from Fig.~\ref{fig:radioimage}. Right: Zoomed in image of the southern lobe and host galaxy from Fig.~\ref{fig:radioimage}. The measured rms noise ($\sigma$) around the source is approximately 68 $\mu$Jy/beam. The contours for both images are at levels of 3, 5, 8, 10, 12, 24, 48 $\times$  $\sigma$. The resolution is 9.4$\times$8.3$\arcsec$ and is shown in the bottom right/left corner as a filled-in ellipse.}%
    \label{fig:closeups}%
\end{figure*}

At a distance of approximately 22\,$\arcmin$ from the field phase centre, a Fanaroff-Riley type II radio galaxy is identified. This radio galaxy is shown in Fig.~\ref{fig:radioimage}, where the lobes of this object are denoted as ``A'' and ``B'' and the central host galaxy, MCG+07-47-10, as ``C''. Closeups of the individual radio lobes are shown in Fig.~\ref{fig:closeups}. Co-ordinates for each of the lobes and the host galaxy are listed in Table~\ref{Tab:locationsandflux}. In particular, lobe ``B'' displays weak extended emission tracing back to the host galaxy (see Fig.~\ref{fig:closeups}, Right). Such weak emission is only now detectable due to the high sensitivity of this observation and would have not been detected in surveys such as the NVSS or FIRST.

\begin{table*}
\centering
\begin{tabular}{lcccccc}
\hline
Src. & ID & Right Ascension & Declination & 1.4\,GHz Flux & 322\,MHz Flux & Spectral\\
      &   &  (J2000)     &   (J2000)  & (mJy) & (mJy) & Index\\
\hline

 A  & Northern Lobe & 23$^\mathrm{h}$ 18$^\mathrm{m}$ 06$^\mathrm{s}$ & +43$\degr$ 18$\arcmin$ 18$\arcsec$ & 7.8$\pm$1.3 & 20.0$\pm$2.0 &  -0.6$\pm$0.3 \\

B & Southern Lobe & 23$^\mathrm{h}$ 18$^\mathrm{m}$ 43$^\mathrm{s}$&  +43$\degr$ 12$\arcmin$ 20$\arcsec$ & 5.5$\pm$0.5 & 16.1$\pm$1.6 & -0.73$\pm$0.21 \\

C & MCG+07-47-10 & 23$^\mathrm{h}$ 18$^\mathrm{m}$ 33$^\mathrm{s}$ &  +43$\degr$ 14$\arcmin$ 49$\arcsec$ &  3.9$\pm$0.4 & 12.4$\pm$1.5 & -0.79$\pm$0.24\\

\hline
\end{tabular}
\caption{Locations of the radio lobes and host galaxy of the spiral DRAGN with integrated fluxes and spectral indices.}
\label{Tab:locationsandflux}
\end{table*}

\begin{figure}[h!]
\centering
\includegraphics[width=0.3\textwidth]{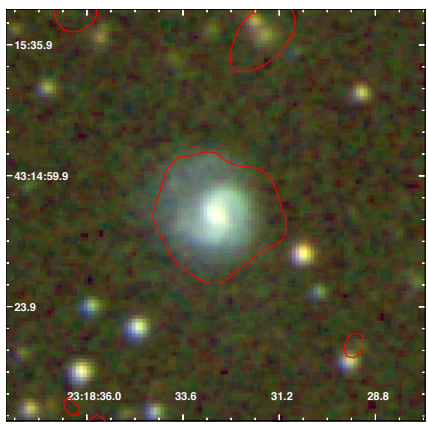}
\caption{Optical DSS image of the host galaxy (MCG+07-47-10) for the spiral DRAGN. The contour shows the 5$\sigma$ level of the 322\,MHz radio data.}
\label{fig:opticalimage}
\end{figure}

On inspection of the Digitized Sky Survey (DSS), the host galaxy (denoted ``C'' in Fig.~\ref{fig:radioimage}), MCG+07-47-10, is a late type galaxy with a classification of Sbc \citep{Voro1968}, a galaxy with spiral arms and a bulge. The limited resolution of the DSS means that the presence of a bar cannot be ruled out. The optical DSS image of this galaxy can be seen in Fig.~\ref{fig:opticalimage} with the 5$\sigma$ radio contour overlaid. The spiral arms of this face-on galaxy are clearly visible and extended radio emission is seen coincident with the optical disk. If this emission was solely due to the core of the AGN, one would expect highly compact emission, unresolved by the GMRT in these data. It can be seen from Fig.~\ref{fig:radioimage} that the radio emission in this case is clearly resolved, with a diameter measured from the 3\,$\sigma$ contour of approximately 45$\arcsec$, equivalent to 5 synthesized beam widths. Such extended radio emission indicates that active star formation is occurring throughout the disk and is due to the injection of Cosmic Ray electrons (CREs) into the interstellar medium (ISM) via the action of Supernovae. These CREs produce non-thermal synchrotron emission, which is observable here.

This area is not covered by the  Sloan Digital Sky Survey \citep[SDSS\footnote{http://www.sdss.org/}; ][]{Alam2015} and currently no spectroscopic redshift information is available for this galaxy. However, MCG+07-47-10 is located close to the galaxy groups NGC\,7618 (z=0.0173) and UGC\,12491 (z=0.0173), the central galaxies of two nearby, roughly equal mass galaxy groups. \cite{Kraft2006} found through X-ray observations that these galaxy groups are in the process of merging. Assuming that MCG+07-47-10 is associated with one of these galaxy groups, we adopt an approximate redshift of $z \approx 0.017$. At this redshift, MCG+07-47-10 would be located at a projected distance of 584 \,kpc (28$\arcmin$)  from NGC\,7618 and 375\,kpc (18$\arcmin$) from UGC\,12491. Given a virial radius for these galaxy groups of approximately 600-800\,kpc \citep{Kraft2006}, MCG+07-47-10 would therefore be found within the virial radius of both groups.

From the GMRT image, we find that this new spiral DRAGN has an angular linear extent of $\sim 9\arcmin$, which equates to a physical size of 207\,kpc at our assumed redshift. Additionally, the optical diameter of MCG+07-47-10 is found to be 12.02\,kpc. 

At an assumed redshift of $z=0.017$ and measured radio flux density values in Table.~\ref{Tab:locationsandflux} indicate a radio luminosity at 322\,MHz for this proposed spiral DRAGN (lobes and host galaxy) of approximately $L_{\rm 322\,MHz}$ = 3.15$\times$10$^{22}$\,Watts\,Hz$^{-1}$ and $L_{\rm 1.4\,GHz}$ = 1.12$\times$10$^{22}$\,Watts\,Hz$^{-1}$. This is significantly weaker to other spiral DRAGNS such as J1649+2635 with $L_{\rm 1.4\,GHz}$=1.03$\times$10$^{24}$\,Watts\,Hz$^{-1}$ \citep{Mao2015}.

\subsection{Additional radio data and integrated fluxes}

MCG+07-47-10 and both lobes are detected by the NVSS \citep{Condon1998}, but not by the VLA Low-frequency Sky Survey redux \citep[VLSSr, ][]{Lane2014,Cohen2007}, TIFR GMRT Sky Survey \citep[TGSS ADR, ][]{Intema2016} nor the WEsterbork Northern Sky Survey \citep[WENSS, ][]{Rengelink1997}. 
In addition, the FIRST survey does not have coverage of this area of the sky.
In the NVSS, the three main components are detected and the catalogue integrated fluxes for each component are shown in Table.~\ref{Tab:locationsandflux}. 

From the combination of the NVSS data and the GMRT data presented in this work, the host galaxy is found to have a spectral index of $\alpha = -0.79 \pm 0.24$, which is a typical value for a star-forming spiral galaxies \citep[$\alpha = -0.74 \pm 0.03$, ][]{Gioia1982}.

\section{Star formation rate}

MCG+07-47-10 is clearly detected in all four WISE bands \citep{Wright2010}. Using colour-colour relationships \citep{Lacy2004,Yan2013}, the measured magnitude difference [W1]-[W2] = 0.184, indicates that MCG+07-47-10 is classified as a star-forming galaxy, with values greater than 0.8 signifying an AGN. Normal spiral galaxies are seen to have a range of magnitude differences of $\approx$ 0.0-0.6 \citep{Wright2010}.

Using the derived relations between star formation rate (SFR) and observed luminosity in the WISE W3 and W4 bands \citep{Jarrett2013},
we find the SFR for MCG+07-47-10 at 12\,$\mu$m to be $\Sigma_{12} = 0.75$\,M$_{\odot}$\,yr$^{-1}$ and at 22\,$\mu$m to be $\Sigma_{22} = 0.51$\,M$_{\odot}$\,yr$^{-1}$. In addition, MCG+07-47-10 is also detected in the near UV by GALEX \citep{Martin2005}. Based on \citet{Kennicutt1998}, we find the UV SFR for MCG+07-47-10 to be $\Sigma_{\rm UV} = 0.16$\,M$_{\odot}$\,yr$^{-1}$. Assuming no contribution from an AGN, the SFR for MCG+07-47-10 derived from the radio data presented here is $\Sigma_{\rm 1.4\,GHz} = 0.63$\,M$_{\odot}$\,yr$^{-1}$ \citep[Eq.21]{Condon1992}, intermediate to the IR and UV derived SFRs.
There are large uncertainties associated with SFR indicators \citep{Hopkins2003}, thus we estimate the SFR to be in the range of
0.16-0.75 \,M$_{\odot}$\,yr$^{-1}$.

The SFR rate in galaxies can show an enormous range from virtually zero in gas poor ellipticals and dwarf-galaxies to  
0.63\,M$_{\odot}$\,yr$^{-1}$ in gas rich spirals \citep{Kennicutt1998}. For comparison MCG+07-47-10 has a similar SFR to the nearby Sb galaxy, NGC2683, with has a SFR $\approx$  0.36\,M$_{\odot}$\,yr$^{-1}$ determined by \citet{Irwin2012}. Compared to other spiral DRAGNS, \citet{Mao2015} found J1649+2635's SFR to be $\approx$ 0.26-2.6 \,M$_{\odot}$\,yr$^{-1}$. The two spiral DRAGNs found by \citet{Singh2015} (J0836+0532,J1159+5820) were found to have 9.99 and 2.89 \,M$_{\odot}$\,yr$^{-1}$ respectively. MCG+07-47-10 with a SFR range of 0.16-0.75 \,M$_{\odot}$\,yr$^{-1}$ would be seen to be in the lower range of the SFR of J1649+2635 and other spiral DRAGNs.

\section{Discussion \& Conclusions}

This paper presents the discovery of the spiral DRAGN MCG+07-47-10. The host galaxy has clearly defined spiral arms, a prominent bulge, and hosts a 188\,kpc DRAGN. 

The radio source, first identified in NVSS, was not previously classified as a DRAGN, but rather three separate radio sources. The deep GMRT observation presented in this paper detects the low surface brightness emission connecting the radio components, thus identifying this radio source as a DRAGN for the first time. 

The central component of the radio emission, from the host galaxy, appears extended in both the NVSS and GMRT data. 
These new GMRT data show resolved radio emission emanating from the entirety of the host galaxy, as opposed to only the core. This suggests that the radio emission is not solely due to the presence of an AGN. Moreover, the total integrated radio flux density for the host galaxy gives a SFR that is in good agreement with SFRs calculated from the IR emission. Resolved radio emission observed throughout the disk suggests that most, if not all the radio emission across the disk is due to star-formation. 

We find that the luminosity of this spiral DRAGN ($L_{\rm 1.4\,GHz}$ = 1.12$\times$10$^{22}$\,Watts\,Hz$^{-1}$) to be significantly lower than other spiral DRAGNS. However, without an reliable redshift this could be misleading. If we take the median redshift from the main galaxy sample from SDSS of $z=0.1$ \citep{Strauss2002}, we would calculate a luminosity of $L_{\rm 1.4\,GHz}$ = 4.25$\times$10$^{23}$\,Watts\,Hz$^{-1}$. This is still lower than that found in other spiral DRAGNS such as J1649+2635 with $L_{\rm 1.4\,GHz}$=1.03$\times$10$^{24}$\,Watts\,Hz$^{-1}$ \citep{Mao2015}. A redshift of 0.1 would imply an optical diameter of 57\,kpc for MCG+07-47-10 and the angular linear extent of the entire spiral DRAGN of $\sim 9$\,$\arcmin$ would equate to a physical size of $\approx$ 1\,Mpc. In order for MCG+07-47-10 to have a comparable luminosity to J1649+2635, MCG+07-47-10 would need to have a redshift of $z \approx 0.17$. This redshift would make the physical size of the DRAGN associated with MCG+07-47-10 significantly larger than 1\,Mpc. We note that this would not necessarily be unusual for the class as two previously identified spiral DRAGNS \citep{Hota2011,Bagchi2014} show evidence of mega-parsec structure. However, the host galaxy, MCG+07-47-10, would then have an optical size of approximately 100\,kpc, which would be exceptionally large compared to similar analogues.

The low-luminosity and low surface brightness of the DRAGN may suggest that the radio emission is old. One possible scenario leading to the formation of this spiral DRAGN is that the host of a low-luminosity DRAGN has had gas injected onto it, perhaps through a merger, and this gas has triggered star-formation and built up spiral arms. Further modelling and observations at several frequencies are required to test this theory.

Assuming that the host galaxy, MCG+07-47-10, is associated with the two galaxy groups in its immediate vicinity (located within the virial radius of NGC\,7618 and UGC\,12491 \citep{Kraft2006}), it would appear to reside in a similarly intermediate density environment to other known spiral DRAGNs \citep{Mao2015}. This supports the idea that spiral DRAGNs require specific environments to form and that a moderately overdense environment is conducive for mergers to trigger a process that transitions ellipticals back to spirals. 

In conclusion, we have presented the discovery of a new spiral DRAGN from observations with the GMRT at 322\,MHz. We have shown that the host galaxy is a star-forming spiral galaxy. Given currently available information, we have demonstrated that the radio luminosity of this new source is significantly lower than that of other spiral DRAGNs, about two orders of magnitude, with a value of $L_{\rm 1.4\,GHz}$ = 1.12$\times$10$^{22}$\,Watts\,Hz$^{-1}$. This may indicate the existence of a previously unknown population of low-luminosity spiral DRAGNs and we suggest that future radio surveys may be used to expand this sample further.

\begin{acknowledgements}
 DM, AMS \& AC gratefully acknowledge support from ERCStG 307215 (LODESTONE). 
 MYM acknowledges support from EC H2020-MSCA-IF-2014 660432 (spiral DRAGNs). 
 We thank the anonymous referee for their helpful comments which improved this paper.

 We acknowledge the usage of the HyperLeda database (http://leda.univ-lyon1.fr) and NASA/IPAC Extragalactic Database.
 The Digitized Sky Surveys were produced at the Space Telescope Science Institute under U.S. Government grant NAG W-2166. The images of these surveys are based on photographic data obtained using the Oschin Schmidt Telescope on Palomar Mountain and the UK Schmidt Telescope. The plates were processed into the present compressed digital form with the permission of these institutions.
\end{acknowledgements}

\bibliographystyle{aa}
\bibliography{ref}

\end{document}